\pdfoutput=1
\documentclass[11pt,twoside,a4paper,cmspaper,final,collab]{cms-tdr}
\def\svnVersion{253bf8f}\def\svnDate{2026/05/27}\def\cmsCernNoTag{CERN-EP-2025-279}\def\cmsCernDate{\today}\def\cmsMessage{Published in Physical Review D as \href{http://dx.doi.org/10.1103/rny9-zby1}{\doi{10.1103/rny9-zby1}.}}
\begin{document}\cmsNoteHeader{HIG-24-011}

\newcommand{\off}{\text{off-shell}\xspace}
\newcommand{\on}{\text{on-shell}\xspace}
\newcommand{\mWW}{\ensuremath{m_{\PW\PW}}\xspace}
\newcommand{\mTH}{\ensuremath{m_{\text{T}}^{\PH}}\xspace}
\newcommand{\GH}{\ensuremath{\Gamma_{\PH}}\xspace}

\cmsNoteHeader{HIG-24-011}

\title{Measurement of the Higgs boson total decay width using the \texorpdfstring{$\PH \to \PW\PW \to \Pe\PGn\PGm\PGn$}{H to WW to e nu mu nu} decay channel in proton-proton collisions at \texorpdfstring{$\sqrt{s} = 13\TeV$}{sqrt(s) = 13 TeV}}

\date{\today}

\abstract{
The Higgs boson (H) decay width is determined from the ratio of off- and on-shell production of $\PH \to \PW\PW \to \Pe\PGn\PGm\PGn$ using proton-proton collision data corresponding to an integrated luminosity of 138\fbinv collected at $\sqrt{s}=13\TeV$ by the CMS experiment at the Large Hadron Collider. The off-shell signal strength is measured as $\mu_{\off} = 1.2^{+0.8}_{-0.7}$. The Higgs boson total decay width is $\GH = 3.9^{+2.7}_{-2.2}$ MeV, in agreement with the standard model prediction. The uncertainty in this result represents a factor of 3 improvement over the previous CMS result in this decay channel.
}

\hypersetup{
pdfauthor={CMS Collaboration},
pdftitle={Measurement of the Higgs boson total decay width using the H to WW to e nu mu nu decay channel in proton-proton collisions at sqrt(s) = 13 TeV},
pdfsubject={CMS},
pdfkeywords={CMS, Higgs, Offshell, HWW}}

\maketitle 

\section{Introduction}\label{sec:intro}
The standard model (SM) is a relativistic quantum field theory that describes the constituents of matter via fields and their associated particle interactions \cite{PhysRevLett.19.1264,GLASHOW1961579,Salam:1968rm}. 
One of the accomplishments of the SM is the prediction of the Higgs field and related Higgs boson (\PH)~\cite{GLASHOW1961579, HIGGS1964132, Higgs1966, PhysRevLett.13.321, PhysRevLett.13.508}, which was discovered in 2012~\cite{ATLAS_2012, CMS_2012}. Since its discovery, a campaign to study its properties has been underway at the CERN Large Hadron Collider (LHC)~\cite{ATLAS:2022vkf, CMSportrait}, a key property being its decay width, $\GH$.

The theoretical prediction of $\GH$ for a SM Higgs boson with a mass of $125\GeV$ is 4.1\MeV with a relative uncertainty of $\pm$3.9\%~\cite{https://doi.org/10.23731/cyrm-2017-002}. Since the value of $\GH$ is much smaller than the invariant mass resolution of the LHC detectors, precise direct measurements of the decay width are not feasible; instead, it can be constrained through the use of off-shell production modes. The CMS Collaboration has previously determined the width to be $\GH = 3.0^{+2.0}_{-1.5}\MeV$~\cite{Hayrapetyan_2025} using off-shell Higgs boson decays to $\PZ\PZ$ with subsequent decays to four charged leptons and to two charged leptons and two neutrinos ($\PH \to \PZ\PZ \to 4\Pell$ and $2\Pell2\PGn$, respectively, with \Pell referring to electrons or muons) using data collected in 2016--2018 at $\sqrt{s}=13\TeV$. The ATLAS Collaboration has reported a width value of $\GH = 4.3^{+2.7}_{-1.9}\MeV$~\cite{ATLAS_width_2025} in the same $\PH \to \PZ\PZ$ decay channels, $\GH = 0.9^{+3.4}_{-0.9}\MeV$ using $\PH \to \PW\PW \to 2\Pell 2\PGn$~\cite{atlascollaboration2025constrainingoffshellhiggsboson}, and $\GH < 450\MeV$ at 95\% confidence level from $\ttbar\ttbar$ production measurements~\cite{Aad_2025}.

To complement previous results, we report the measurement of $\GH$ from the ratio of its off- and on-shell production in the $\PH \to \PW\PW \to \Pe\PGn\PGm\PGn$ decay mode, using CMS data collected in 2016--2018 at $\sqrt{s}=13\TeV$. In this paper, the off-shell contribution is defined using a requirement on the $\PW\PW$ invariant mass at the generator level of $\mWW > 160\GeV$. The previous CMS result using this channel, $\GH < 26\MeV$~\cite{CMS_HWWwidth_Run1}, was performed with $\sqrt{s} = 7$ and $8\TeV$ data collected in 2011--2012. The increased center-of-mass energy and integrated luminosity, together with reduced experimental and theoretical uncertainties, motivate this study. Deviations of $\GH$ from its SM expectation could point to physics beyond the SM, such as anomalous couplings, since the decay width is determined by all known particle couplings to the Higgs boson.

Given that the width of the Higgs boson is narrow compared to its mass, the narrow-width approximation can be applied \cite{Kauer:2012hd, PhysRevD.37.1801, Dixon_2013, Campbell_2011}, and its on-shell production cross section $\sigma_{\on}$ reduces to:
\begin{equation}\label{eq:prop}
    \sigma_{\on} \approx \frac{g_{\text{p}}^2g_{\text{d}}^2}{(p^2-M_{\PH}^2) +M_{\PH}\GH} \approx \frac{g^2_{\text{p}}g^2_{\text{d}}}{M_{\PH}\GH},
\end{equation}
where $p$ is the four-momentum of the Higgs boson, $M_{\PH}$ its mass, and $g_{\text{p}}$ and $g_{\text{d}}$ the respective production and decay mode couplings being considered. At center-of-mass energies well above the $\PW\PW$ production threshold, there is no significant dependence of the corresponding off-shell cross section $\sigma_{\off}$ on $\GH$ and $M_{\PH}$, so we simply have:
\begin{align}
    \mu_{\off} \propto \sigma_{\off} = g^2_{\text{p}} g^2_{\text{d}}, \\
    \mu_{\on} \propto \sigma_{\on} \propto \frac{g^2_{\text{p}} g^2_{\text{d}}}{\GH},
\end{align}
where $\mu$ is the signal strength, defined as the ratio of the measured cross section to the SM prediction for either on- or off-shell production, both of which are equal to unity in the SM. Hence,
\begin{equation}
\GH\propto\frac{\mu_{\off}}{\mu_{\on}}.
\end{equation}\label{eq:width_intro}
Thus, it is possible to determine the decay width of the Higgs boson by measuring the relative rates of off- to on-shell Higgs boson production.

\begin{figure*}[htbp]
    \centering
        \includegraphics[width=0.35\textwidth]{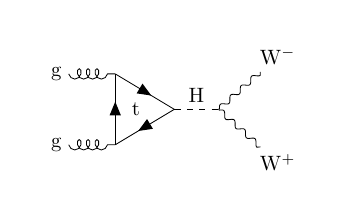}
        \includegraphics[width=0.35\textwidth]{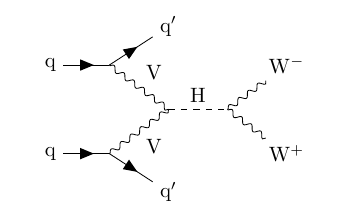}
        \caption{Feynman diagrams illustrating Higgs boson production and decays to $\PW\PW$ in ggF (left) and VBF (right) modes, where H refers to an on- or off-shell Higgs boson.}
        \label{fig:higgs_production}
  \end{figure*}

The two largest Higgs boson production modes, gluon-gluon fusion (ggF) and vector boson fusion (VBF, where $\PV = \PW$ or \PZ), illustrated by the Feynman diagrams in Fig.~\ref{fig:higgs_production}, are used in this analysis. The $\PW\PW \to \Pe\PGn\PGm\PGn$ final state is chosen, where the different lepton flavors are required to remove the large Drell--Yan (DY) background in the same-flavor final state. Neutrinos present a challenge because they are electrically neutral, weakly interacting particles, which makes them impossible to directly detect. Their presence can only be inferred from missing transverse momentum, \ptvecmiss, as defined in Section~\ref{sec:detector}.

There are several background physics processes that contribute to the $\PH \to \PW\PW$ channel. At large $\mWW$, the most relevant ones for this measurement are the gluon-induced continuum $\Pg\Pg \to \PW\PW$ and electroweak $\Pq\Pq \to \Pq\Pq\PW\PW$ production, which interfere destructively with off-shell ggF and VBF, respectively, because of the identical initial and final states. The destructive interference prevents the cross section of scattered \PW bosons from diverging to infinity, thereby preserving unitarity in the electroweak theory. In this paper, the off-shell contribution is defined as the signal, seen in Fig.~\ref{fig:higgs_production}, plus interference, and it accounts for only a few percent of the total expected event yield once all backgrounds are included, especially the on-shell contributions, \ttbar, and nonresonant $\PW\PW$ production. As a result, the analysis effectively searches for a small shape distortion in the high-mass tail of the $\mWW$ distribution on top of sizeable backgrounds, and relies on a precise modeling of off-shell $\PH \to \PW\PW$, the nonresonant $\PW\PW$ continuum, and their interference.

The structure of this paper is as follows. Section~\ref{sec:detector} provides a short overview of the CMS detector. Details of the data samples and the simulated events are presented in Section~\ref{sec:data_and_sim}. 
The strategy for event selection and the categorization procedure are outlined in Section~\ref{sec:eventsel}.  Section~\ref{sec:bkgmodel} discusses the background estimation, while Section~\ref{sec:unc} summarizes the treatment of systematic uncertainties. The statistical procedure and final results are reported in Sections~\ref{sec:Combine} and~\ref{sec:results}, respectively, and the final summary is given in Section~\ref{sec:conclusion}. Tabulated results are provided in the HEPData record for this analysis~\cite{hepdata}.

\section{The CMS detector and event reconstruction}\label{sec:detector}
The CMS apparatus~\cite{CMS:2008xjf,CMS:2023gfb} is a multipurpose, nearly hermetic detector, designed to trigger on~\cite{CMS:2020cmk,CMS:2016ngn,Hayrapetyan_2024} and identify electrons, muons, photons, and (charged and neutral) hadrons~\cite{cms_egm_run2,muon_reco,CMS:2014pgm}. Its central feature is a superconducting solenoid of 6\unit{m} internal diameter, providing a magnetic field of 3.8\unit{T}. Within the solenoid volume are a silicon pixel and strip tracker, a lead tungstate crystal electromagnetic calorimeter (ECAL), and a brass and scintillator hadron calorimeter (HCAL), each composed of a barrel and two endcap sections. Forward calorimeters extend the pseudorapidity $\eta$ coverage provided by the barrel and endcap detectors. Muons are reconstructed using gas-ionization detectors embedded in the steel flux-return yoke outside the solenoid. More detailed descriptions of the CMS detector, together with a definition of the coordinate system used and the relevant kinematic variables, can be found in Refs.~\cite{CMS:2008xjf,CMS:2023gfb}.

Events of interest are selected using a two-tiered trigger system. The first level (L1), composed of custom hardware processors, uses information from the calorimeters and muon detectors to select events at a rate of around 100\unit{kHz} with a fixed latency of 4\mus~\cite{CMS:2020cmk}. The second level, known as the high-level trigger (HLT), consists of a farm of processors running a version of the full event reconstruction software optimized for fast processing, and reduces the event rate to a few kHz before data storage~\cite{CMS:2016ngn,Hayrapetyan_2024}.

The particle-flow (PF) global event reconstruction~\cite{CMS:2017yfk} combines information from the tracking system, calorimeters, and muon detectors to reconstruct and identify individual particles (photons, electrons, muons, charged and neutral hadrons). Electrons are reconstructed from ECAL energy clusters matched to tracks in the silicon tracker, with an explicit treatment of bremsstrahlung along the trajectory, and are accepted within $\abs{\eta}<2.5$. Muons are reconstructed by matching tracks in the silicon tracker to compatible hits or segments in the muon system within $\abs{\eta}<2.4$. Tracks not identified as electrons or muons are assigned to charged hadrons, while neutral hadrons are reconstructed from HCAL energy deposits not associated with any charged particle track, or from a combined ECAL+HCAL excess over the expected charged-hadron response. 

The energy of electrons is determined from a combination of the track momentum at the main interaction vertex, the corresponding ECAL cluster energy, and the energy sum of all bremsstrahlung photons attached to the track. The momentum (energy) of muons is obtained from the corresponding curvature of the track. Only the sum of the transverse momentum (\pt) carried away by the neutrinos can be reconstructed using missing transverse momentum. The missing transverse momentum vector \ptvecmiss is computed as the negative vector sum of the transverse momenta of all the PF candidates in an event, and its magnitude is denoted as \ptmiss~\cite{cms_met_puppi_run2}. Additional proton-proton interactions within the same or nearby bunch crossings (pileup) can contribute additional tracks and calorimetric energy depositions to the jet momentum. To mitigate this effect, charged particles identified to be originating from pileup vertices are discarded and an offset correction~\cite{cms_pileup_mitigation_run2} is applied to correct for remaining contributions. The value of \ptmiss is defined using the pileup per particle identification (PUPPI) algorithm~\cite{cms_met_puppi_run2} along with specific filters to mitigate the effects of the detector noise, and is required to exceed 20\GeV. Anomalous high-\ptmiss events can be due to a variety of reconstruction failures, detector malfunctions, or noncollision backgrounds. Such events are rejected by event filters that are designed to identify more than 85--90\% of the spurious high-\ptmiss events with a mistagging rate less than 0.1\%~\cite{cms_met_puppi_run2}.

For each event, hadronic jets with a minimum \pt of 20\GeV are clustered from reconstructed particles using the infrared and collinear safe anti-\kt algorithm~\cite{Cacciari:2008gp, Cacciari:2011ma} with a distance parameter of 0.4, within $\abs{\eta}<4.7$. Jet momentum is determined as the vectorial sum of all particle momenta in the jet, and is found from simulation to be, on average, within 5--10\% of the true momentum over the entire \pt spectrum and detector acceptance. 
Jet energy corrections are derived from simulation to bring the measured response of jets to that of particle-level jets on average. In situ measurements of the momentum balance in multijet events are used to account for any residual differences in the jet energy scale between data and simulation. The jet energy resolution amounts typically to 15--20\% at 30\GeV, 10\% at 100\GeV, and 5\% at 1\TeV~\cite{CMS:2016lmd}. Additional selection criteria~\cite{cms_pileup_mitigation_run2} are applied to each jet to remove jets potentially dominated by anomalous contributions from various subdetector components or reconstruction failures.

\section{Data and simulated samples}\label{sec:data_and_sim}
In this analysis, we use proton-proton collision data samples collected by the CMS experiment in 2016--2018, with a total integrated luminosity of 138\fbinv~\cite{CMS:2021xjt,CMS:2018elu,CMS:2019jhq}.

Each event is required to pass HLT selections, which involve reconstruction of leptons (\Pe, \PGm) passing specific isolation and identification criteria \cite{Hayrapetyan_2024}. The single-lepton trigger selects either one electron with $\pt > 25$, $35$, and $32\GeV$ for the 2016, 2017, and 2018 datasets, respectively, or one muon with $\pt > 24\,(27)\GeV$ for the 2016 and 2018 (2017) datasets. For the combined electron-muon trigger, the highest transverse momentum (leading) lepton must have $\pt > 23\GeV$ with the requisite second-highest transverse momentum (subleading) lepton having $\pt > 12\GeV$. The average trigger efficiency, stable across all years to within about 1\%, for a single (double) lepton trigger is 87 (95)\%~\cite{Hayrapetyan_2024}.                                               

Simulated events are used for modeling signal and background processes. Event samples for the 2016, 2017, and 2018 datasets are simulated separately to account for differences in the detector and pileup conditions.

\begin{figure*}[htbp]
\centering
        \includegraphics[width=0.35\textwidth]{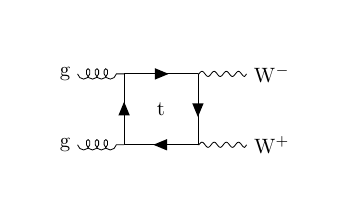}
        \includegraphics[width=0.35\textwidth]{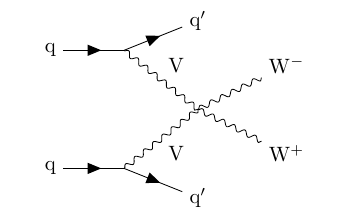}
        \caption{Feynman diagrams for nonresonant $\PW\PW$ production: $\Pg\Pg \to \PW\PW$ (left) and one example of $\Pq\Pq \to \Pq\Pq\PW\PW$ (right).}
        \label{fig:ggWW}
\end{figure*}

The VBF off-shell Higgs boson simulations are handled following the procedure described in Ref.~\cite{CMSNote}, which combines several Higgs boson samples to reproduce the expected line shape for signal ($S$), the $\Pq\Pq \to \Pq\Pq\PW\PW$ continuum background ($B$), shown in Fig.~\ref{fig:ggWW}, and the combination of signal, background, and interference ($S+B+I$). Higgs boson production is generated with \POWHEG~v2~\cite{Nason:2004rx,Frixione_2007,Alioli:2010xd} at next-to-leading order (NLO), while the decay to a pair of \PW bosons and subsequently to leptons is performed using \textsc{JHUgen}~v7.1.4~\cite{Bolognesi:2012mm}. The VBF samples are generated for $\mWW$ values ranging from 125--3000\GeV, with sufficient event counts to cover both the on- and off-shell regions. The simulated Higgs bosons are treated within the complex pole scheme (CPS) \cite{Goria:2011wa}, and reweighted to the SM expectation at $M_{\PH} = 125\GeV$ using the matrix-element likelihood approach (MELA) \cite{Gritsan:2016hjl,Gritsan2020}. These samples are combined, resulting in a consistent set of off-shell VBF samples with high number of entries per mass bin in the off-shell tail.

Off-shell ggF samples are generated using the \MCFM~v7.0~\cite{MCFMCampbell:1999ah,MCFMCampbell:2011bn,MCFMCampbell:2015qma} generator, which can handle leading-order (LO) generation for the ggF $S$, the $\Pg\Pg \to \PW\PW$ $B$, and $S+B+I$. The use of \MCFM in this analysis differs from the procedure recommended in Ref.~\cite{CMSNote} because matrix-element reweighting of quark-mediated box diagrams, such as those for the $\Pg\Pg \to \PW\PW$ background, illustrated in Fig.~\ref{fig:ggWW}, yields divergent results. The \MCFM generator does not use the reweighting procedure and naturally accounts for the impact of the top quark loop on the Higgs boson amplitude. Off-shell ggF samples are first normalized to next-to-NLO (NNLO) precision via a K-factor~\cite{CMS:coupling}, differential in \mWW, and then normalized at next-to-NNLO ($\text{N}^3\text{LO}$)~\cite{Anastasiou:2016cez, https://doi.org/10.23731/cyrm-2017-002}.

A requirement of $\mWW > 160\GeV$ is set at the generator level to define the ggF and VBF off-shell events in the simulated samples, as this value corresponds to the threshold above which the Higgs boson can decay into two on-shell W bosons. The selection is inverted for on-shell events, where at least one W boson is off-shell. Generated events where at least one $\PW$ boson decays into a $\PGt$ lepton, which subsequently decays leptonically, are treated as signal. These leptonic $\PGt$ decays contribute approximately 3--5\% to the signal yield in the on-shell regions and 10--15\% in the off-shell regions.

The ggF and VBF on-shell Higgs boson samples are simulated using \POWHEG$+$\textsc{JHUgen}. The ggF events are reweighted to match the parton shower (PS) model at NNLO+PS~\cite{Hamilton:2013fea,Hamilton:2015nsa}, and further scaled with a K-factor to $\text{N}^3\text{LO}$~\cite{Anastasiou:2016cez, https://doi.org/10.23731/cyrm-2017-002}. Simulated samples for the associated production (mainly \PW{}\PH and \ttbar{}\PH) are also considered, as described in Ref.~\cite{HWW_xsec_Run2}, although their contribution in the analysis phase space is negligible.
The background processes are simulated using several different generators.
The production of $\PWp\PWm$ pairs via quark-antiquark annihilation, $\Pq\Pq \to \PW\PW$, is simulated at NNLO+PS with \POWHEG.
For this sample, a discrepancy between data and simulation is observed in the $\pt^{\PW\PW}$ distribution. This particular kinematic variable is difficult to model for $\PW\PW$ processes, especially at low $\pt^{\PW\PW}$ values around 1\GeV. To improve the Monte Carlo (MC) modeling, a correction factor is derived in the inclusive phase space for $\PW\PW$ production. After subtracting all non-$\Pq\Pq \to \PW\PW$ backgrounds from the data, the ratio of the resulting $\pt^{\PW\PW}$ distribution to the $\Pq\Pq \to \PW\PW$ MC prediction is fitted with a second-order polynomial, which is then applied as an event weight. The overall $\Pq\Pq \to \PW\PW$ normalization before and after applying the correction is preserved. The DY production of $\PZ/\PGg^{*}$, and multiboson processes, such as $\PW\PGg$, $\PZ\PZ$, and $\PV\PV\PV$, are generated at NLO using \MGvATNLO~v2.4.2~\cite{Alwall_2014}. Single top quark, \ttbar, $\PW\PZ$, and $\PW\PGg^*$ background processes are simulated at NLO with \POWHEG~v2. The measured top quark \pt distribution from \ttbar events in data is softer~\cite{softtop} than that predicted by simulation. This discrepancy is due to missing higher-order quantum chromodynamic (QCD) and electroweak corrections. A paramaterized weight correction~\cite{topcorrCzakon_2017} is applied in simulation as a function of the generator-level top quark \pt distribution ratio (from NNLO calculations and NLO samples) to improve the modeling in the NLO \ttbar MC simulation. 

The $\PYTHIA8$ generator~\cite{Sjostrand:2014zea} with the CP5 underlying event (UE) tune~\cite{Sirunyan:2019dfx} is used for PS, hadronization, and UE simulation. Parton distribution functions (PDFs) are modeled with the NNPDF3.1~\cite{Ball:2017nwa} set at NNLO precision for all samples. For all processes, the detector response is simulated using a detailed description of the CMS detector, based on the \GEANTfour toolkit~\cite{Agostinelli:2002hh}.  The distribution of the number of pileup interactions in the simulation is adjusted to match that observed in data for each year.
Additional corrections are applied to the MC simulated samples to match the experimental response in data for the trigger selection, lepton reconstruction, identification, isolation, and momentum scale, jet energy scale and resolution, missing transverse energy scale, \PQb tagging selection, and L1 trigger corrections for events that are identified with an incorrect bunch crossing~\cite{CMS:2020cmk} (a ``prefiring'' correction).

\section{Event selection and categorization}\label{sec:eventsel}
The dilepton objects are defined as the oppositely charged electron and muon pair, passing $\pt > 25\,(20)\GeV$ for the leading (subleading) lepton.
Several quality requirements are applied to define a muon. We use tight identification and isolation criteria~\cite{muon_reco}. Additional selections on the distance from the primary interaction vertex (PV) in the transverse and longitudinal planes, $d_{xy} < 0.02\unit{cm}$ and $d_z < 0.1\unit{cm}$, are required to suppress muons from pileup or from heavy-quark hadron decays (nonprompt).
To reduce the misidentification rates due to nonprompt leptons not coming from the PV, a multivariate (MVA) discriminant~\cite{cms_tth_multilepton_2018} is applied to ensure a maximum separation between prompt and nonprompt leptons.
Electrons are identified in a similar fashion, based on a multivariate identification and tight isolation criteria~\cite{cms_egm_run2, HWW_xsec_Run2}. The transverse and longitudinal impact parameter requirements are: $d_{xy} < 0.05\,(0.10)\unit{cm}$ and $d_z < 0.1\,(0.2)\unit{cm}$ for leptons in the pseudorapidity range of the barrel (endcap).

The first selection criterion from the $\PH \to \PW\PW \to \Pe\PGn\PGm\PGn$ decay chain requires $\ptmiss > 20\GeV$ to account for the undetected neutrinos. The dileptons are then required to have an invariant mass $m_{\Pe\PGm} > 20\GeV$ to suppress nonprompt lepton pairs, and a total transverse momentum magnitude $\pt^{\Pe\PGm} > 30\GeV$ to reduce the contributions from $\PZ / \gamma^{*} \to \PGtp\PGtm$ and events in which one of the leptons is nonprompt. The reconstructed transverse mass of the Higgs boson is defined as:
\begin{equation}
\mTH = \sqrt{2\pt^{\Pe\PGm} \ptmiss (1-\cos \Delta \phi)},
\end{equation}
where $\Delta\phi$ is the difference in azimuthal angle between the dilepton \ptvec and \ptvecmiss. To further suppress the DY background, we require $\mTH > 60\GeV$. In order to reduce background from top quark production, the \textsc{DeepJet} \PQb tagging algorithm~\cite{Sirunyan:2017ezt, Bols:2020bkb, CMS-DP-2023-005} is used on jets of $\pt > 20\GeV$ to identify (\PQb tag) those jets, labeled \PQb jets, coming from the hadronization of \PQb quarks. A working point with a misidentification (mistag) rate of 1\% and a \PQb jet efficiency of $\approx\,$80\% versus light-quark-composite jets is selected per year. Events are required to have no \PQb tagged jets.

Events are first categorized based on the number of jets present, using an algorithm that has tight identification and loose pileup mitigation criteria~\cite{cms_pileup_mitigation_run2}, in three different bins: 0, 1, and $\ge$2 jets, where all jets must have $\pt > 30\GeV$ to be counted. Events from ggF populate mostly the 0- or 1- jet categories, with the jets coming from QCD radiation, whereas VBF events have at least two high-\pt jets at large pseudorapidities. 

To better discriminate between signal and background processes and further categorize the events, we used a multiclass fully connected deep neural network (DNN) trained with the \textsc{Keras} library~\cite{Chollet2021DeepLearning} for each of the three jet categories. Four classes are considered in both the 0- and 1-jet categories: off-shell ggF, on-shell ggF, top quark production, and nonresonant $\PW\PW$ production, corresponding to the one signal region (SR) and the three largest control regions (CRs), respectively. Six classes are defined for the $\geq$2-jet category: off-shell ggF, off-shell VBF, on-shell ggF, on-shell VBF, top quark production, and nonresonant $\PW\PW$ production, corresponding to the two SRs and four CRs.

The DNN inputs and event selection depend on the number of jets in each event. The first-layer DNN variables for the $\geq$2-jet category are defined as follows. For the dileptons, the variables are defined as their invariant mass, the differences in the pseudorapidity ($\Delta\eta$) and azimuthal angle ($\Delta\phi$), $\Delta R$ (defined as $\sqrt{\Delta\eta^2+\Delta\phi^2}$), the total \pt of the two leptons, and the \pt of the leading and subleading leptons. For dijets, we use the reconstructed transverse mass, $\Delta\eta$, $\Delta\phi$, and the \pt and \PQb tagging score of the leading and subleading jets. The \ptmiss and the difference in azimuthal angle $\Delta\phi$ between $\ptvec^{\Pe\PGm}$ and \ptvecmiss are used as variables to detect the presence of neutrinos in the event. The Higgs boson transverse mass \mTH is also included as an input variable. Finally, a proxy invariant mass of the $\PW\PW$ system is constructed assuming that the neutrinos from the \PW boson decays travel in the same direction as their corresponding charged leptons. To compute this variable, \ptvecmiss is projected onto the azimuthal direction of each reconstructed lepton. This projected momentum is then added to the transverse momentum of each lepton to form the total transverse momentum of the lepton and neutrino system, whose mass is set to the nominal \PW boson mass. The final variable is defined as the invariant mass of the sum of these two reconstructed \PW boson candidates.

The number of nodes in successive layers of the DNN goes from 16 to 60, and then decreases to 30, 15, and 10. This produces the final six event categories: off-shell ggF and VBF, on-shell ggF and VBF, top quark production, and nonresonant $\PW\PW$ production.
The 0- and 1-jet DNNs are trained separately and are similar to the $\geq$2-jet DNN, with the exception of non-jet or single-jet variables.

The DNN is trained with the Adam optimizer \cite{DNNkingma2017adammethodstochasticoptimization}, and the output vector is converted into probability distributions via the softmax function, such that the sum of the scores of all classes adds to unity. Events from all MC samples of the processes considered are randomly split such that $50\%$ of the events are used for testing after training, $17.5\%$ for validation during training, and $32.5\%$ for the training itself. To avoid overtraining, an early stopping mechanism is employed that truncates the training if the validation loss submits the same result within 30 epochs. Once trained, a Kolmogorov--Smirnov (KS)~\cite{KS} test is performed comparing the DNN output distributions for the training and testing datasets. The test determined that no bias was introduced in the training set.

Analysis SRs and CRs are defined by requiring the output DNN score for each training class to be above $0.5$. For example, in the off-shell ggF SR category, we demand the corresponding $\text{DNN}_{\off} > 0.5$. The nonresonant $\PW\PW$ and \ttbar CRs have the same requirement, along with an additional $\text{DNN}_{\text{top}} < 0.2$ or an inverted \PQb quark veto selection, respectively.

\section{Background estimation}\label{sec:bkgmodel}
All background contributions, with the exception of those arising from nonprompt leptons, are estimated using MC simulations. 

The nonprompt lepton background is determined directly from data~\cite{HWW_xsec_Run2} by analyzing events in which the lepton identification criteria are relaxed.  Such loosely identified leptons are more likely to originate from jets that have been misclassified as isolated leptons. By evaluating the probability that the leptons selected with these loose criteria pass the SR requirements, together with the efficiency for genuine leptons to be reconstructed and identified, the yield of nonprompt leptons in the SR can be inferred~\cite{HWW_xsec_Run2}. The reliability of the procedure is verified in a dedicated validation region by requiring a same-sign \Pe-\PGm pair, while preserving the remaining SR selections. 

The overall normalizations of the two main background processes, the nonresonant $\PW\PW$ and \ttbar, are constrained from data simultaneously in the SRs and in the dedicated CRs.

\section{Uncertainties}\label{sec:unc}
All the systematic uncertainties described in Ref.~\cite{HWW_xsec_Run2} have been considered in this analysis.
The experimental systematic uncertainties are estimated from the correlated and uncorrelated integrated luminosity uncertainties; the trigger and lepton isolation and identification efficiencies from the differences between data and MC simulation; the jet energy scale and resolution from variations due to correlated and uncorrelated parameters; the unclustered energy scale; the \PQb tagging selection efficiency using scale factors; the pileup correction; and the MC modeling uncertainties from nonprompt, nonresonant $\PW\PW$, and top quark backgrounds. The nonprompt lepton background is influenced by the jet flavor composition when jets are misidentified as leptons. Such effects contribute an uncertainty of about 5--10\% to the misidentification rate~\cite{Sirunyan:2018egh}. An additional normalization uncertainty of 30\% is assigned to the nonprompt lepton background to cover the maximum discrepancy between data and MC simulation in the same-sign lepton validation region. As a conservative approach, uncertainties equal to the full top quark \pt correction for \ttbar events and the full $\pt^{\PW\PW}$ correction for nonresonant $\PW\PW$ events are applied.

The theoretical systematic uncertainties arise in the choices of PDFs; renormalization and factorization scales; PS modeling of jet multiplicity; and UE modeling. 

Each systematic uncertainty is incorporated into the statistical model using a nuisance parameter, which can either modify the shape or perturb the scale of the DNN output used to derive the final decay width. All of the nuisance parameters are symmetric Gaussian constraints. The contribution of each overall type of uncertainty to the final measurements is given in the summary table of Section~\ref{sec:results}. The largest systematic uncertainty is from the theoretical components, with the normalization and factorization uncertainties being the most important.

\section{Statistical procedure}\label{sec:Combine}
To extract the off-shell signal strength, $\mu_{\off}$, and the Higgs boson total decay width, $\GH$, the CMS statistical analysis tool \textsc{Combine}~\cite{combine_2024} is used.
This framework is based on \textsc{RooFit}~\cite{Verkerke:2003ir} and \textsc{RooStats}~\cite{Moneta:2010pm}. Using \textsc{Combine}, we build the likelihood function and perform scans to extract the analysis results. 
The measurements are made from a profile maximum likelihood fit to different observables depending on the analysis categories: the DNN output distribution is used for the off-shell SRs ($\text{DNN}_{\off}$) and on-shell CRs ($\text{DNN}_{\on}$), while the event yields in the $\PW\PW$ and \ttbar CRs are accounted for by their normalization factors. In the 0- and 1-jet categories, we include the ggF off-shell SR, ggF on-shell CR, and \ttbar CR. The $\geq$2-jet category includes the VBF off-shell SR, VBF on-shell CR, and $\PW\PW$ CR, in addition to the other three. Thus, a total of 12 regions are combined per year.

The likelihood fits within \textsc{Combine} take probability density functions as inputs, which are required to be positive-definite. Because the interference contribution can take negative values, the inputs are redefined: signal ($S$), background ($B$), and the combined signal + background + interference ($S{+}B{+}I$). This ensures positive-definite templates while preserving the correct dependence on the off-shell signal strength. Rewriting the total event yield in terms of the $S$, $B$, and $I$ inputs,
\ifthenelse{\boolean{cms@external}}
{
\begin{multline*}
 Y = \sqrt{\mu_{\off}} (S+B+I) + (\mu_{\off} - \sqrt{\mu_{\off}})(S) \\ + (1 - \sqrt{\mu_{\off}})(B), 
\end{multline*}
}
{
\begin{equation*}
 Y = \sqrt{\mu_{\off}} (S+B+I) + (\mu_{\off} - \sqrt{\mu_{\off}})(S) + (1 - \sqrt{\mu_{\off}})(B),
\end{equation*}
}
where $\mu_{\off}$ is the off-shell signal strength, and each template, \eg, ($S{+}B{+}I$), is its own input. Maximizing the likelihood function then gives us a measurement of $\mu_{\off}$.
When extracting the value of $\GH$, a similar procedure is followed.
Instead of a direct observation of the width, the ratio of off- and on-shell signal strengths is used:
\begin{equation}
  \label{eq:width}
  \text{r} = \frac{\GH}{\GH^{\text{SM}}} = \frac{\mu_{\off}}{\mu_{\on}}.
\end{equation}
The ratio r is treated as the parameter of interest, and the yield equation is rewritten using the relationship between r and $\mu_{\off}$, giving:
\ifthenelse{\boolean{cms@external}}
{
\begin{multline*}
 Y = \sqrt{\text{r}\mu_{\on}} (S+B+I) + (\text{r}\mu_{\on} - \sqrt{\text{r}\mu_{\on}})(S) \\ + (1 - \sqrt{\text{r}\mu_{\on}})(B),
\end{multline*}
}
{
\begin{equation*}
 Y = \sqrt{\text{r}\mu_{\on}} (S+B+I) + (\text{r}\mu_{\on} - \sqrt{\text{r}\mu_{\on}})(S) + (1 - \sqrt{\text{r}\mu_{\on}})(B),
\end{equation*}
}
where $\mu_{\on}$ is determined from the on-shell CRs. The contamination of on-shell (off-shell) in off-shell (on-shell) regions is negligible, so that we decouple r and $\mu_{\on}$.

\begin{figure*}[htbp]
  \centering
  \includegraphics[width=0.49\textwidth]{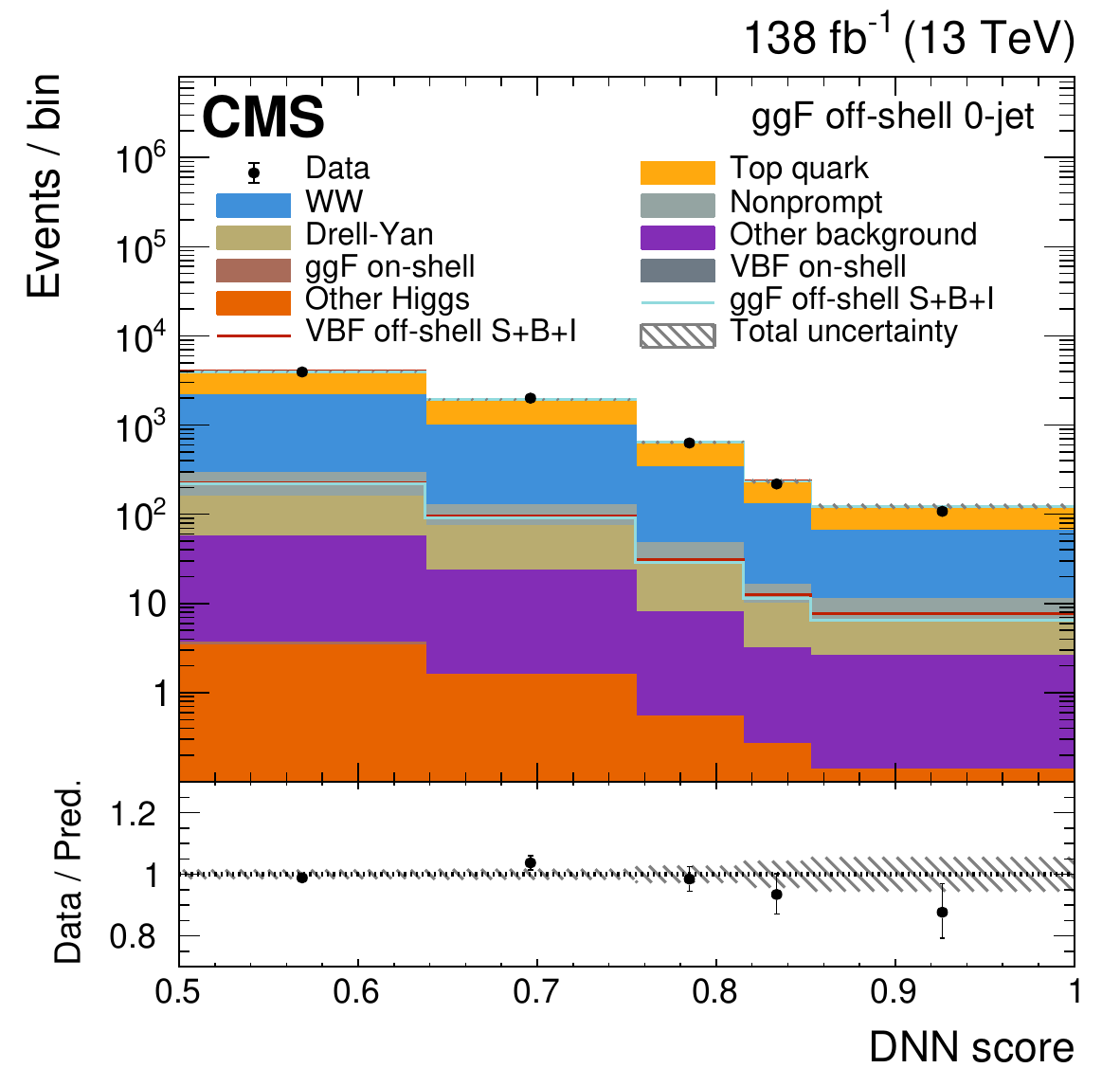}
  \includegraphics[width=0.49\textwidth]{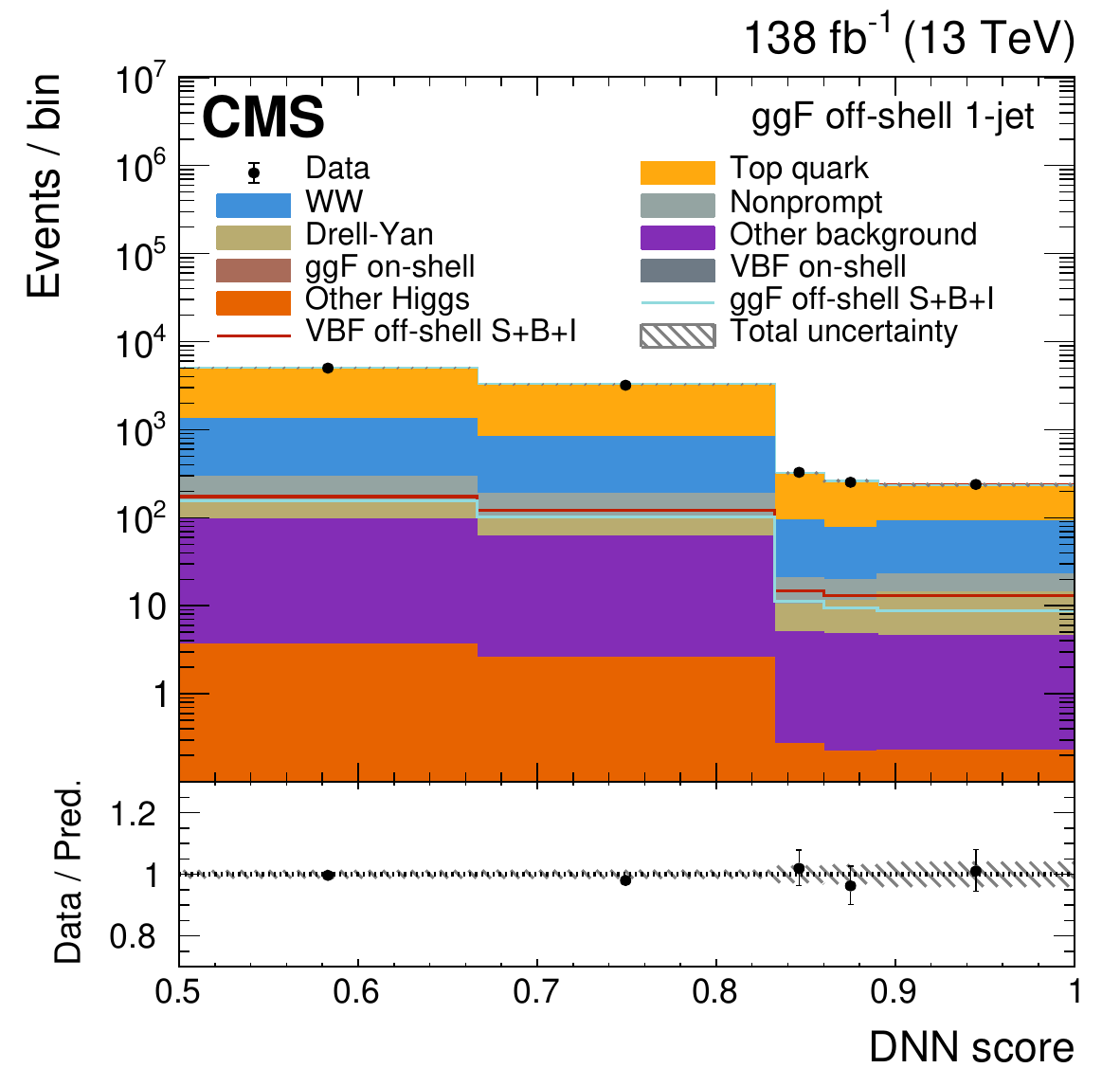} \\
  \includegraphics[width=0.49\textwidth]{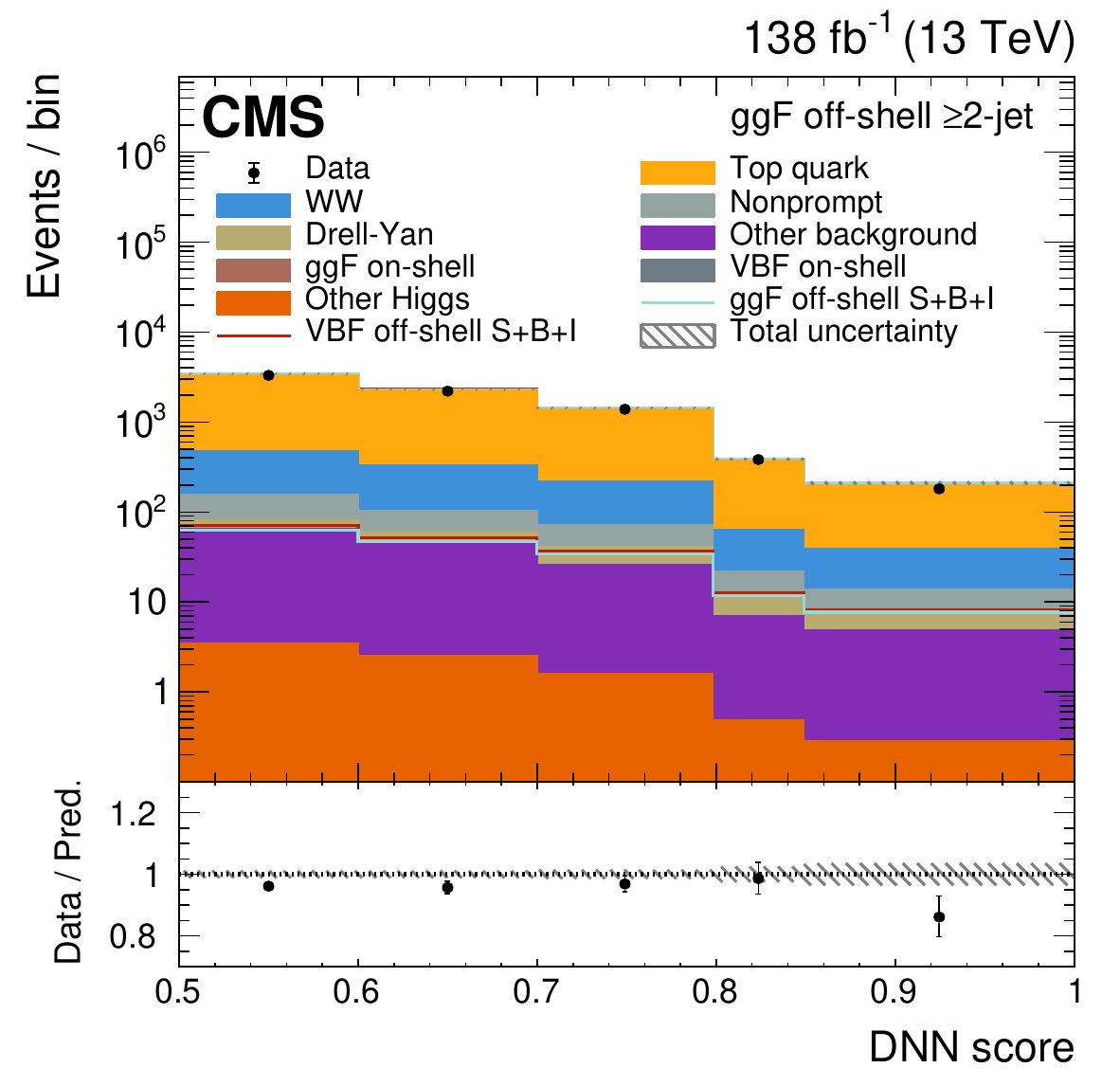}
  \includegraphics[width=0.49\textwidth]{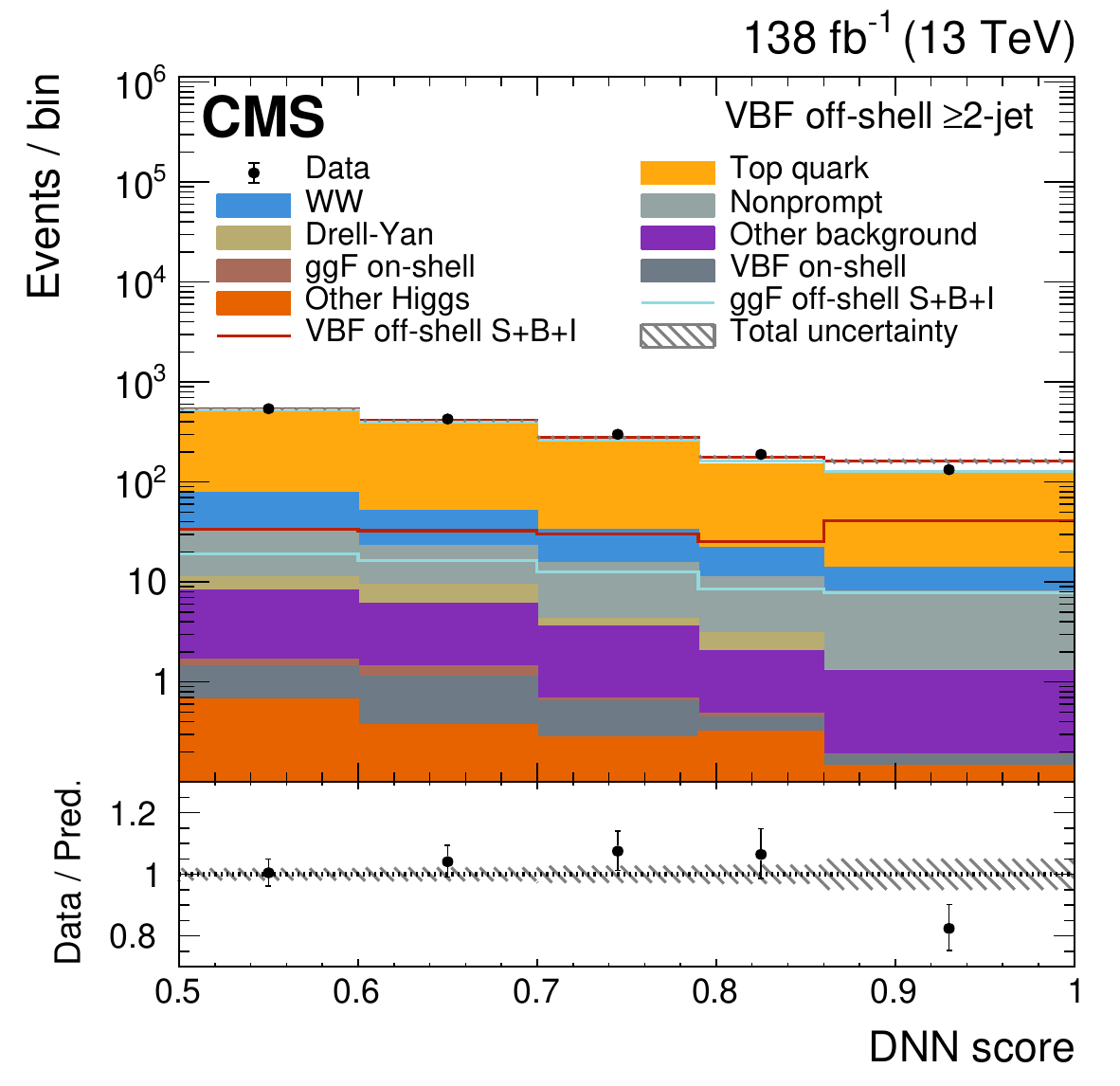}
  \caption{The off-shell SR post-fit DNN output distributions for the 0-jet (upper left), 1-jet (upper right), and $\geq$2-jet (lower plots) event categories. The data are shown as the black dots, with the vertical bars representing the statistical uncertainties. The predicted signal and background distributions are shown by the various colored histograms. The predicted signal ggF and VBF off-shell $S+B+I$ distributions are displayed both as part of the total stacked histograms, along with all the predicted background distributions, and on top of the stacked background distributions for visibility. The lower portion of the plots gives the ratio of the data to the total predicted yields, including the post-fit signal. The hatched area represents the total uncertainty in the ratio.}
  \label{fig:post-fit_dnnScores}
\end{figure*}

\begin{figure*}[htbp]
  \centering
  \includegraphics[width=0.49\textwidth]{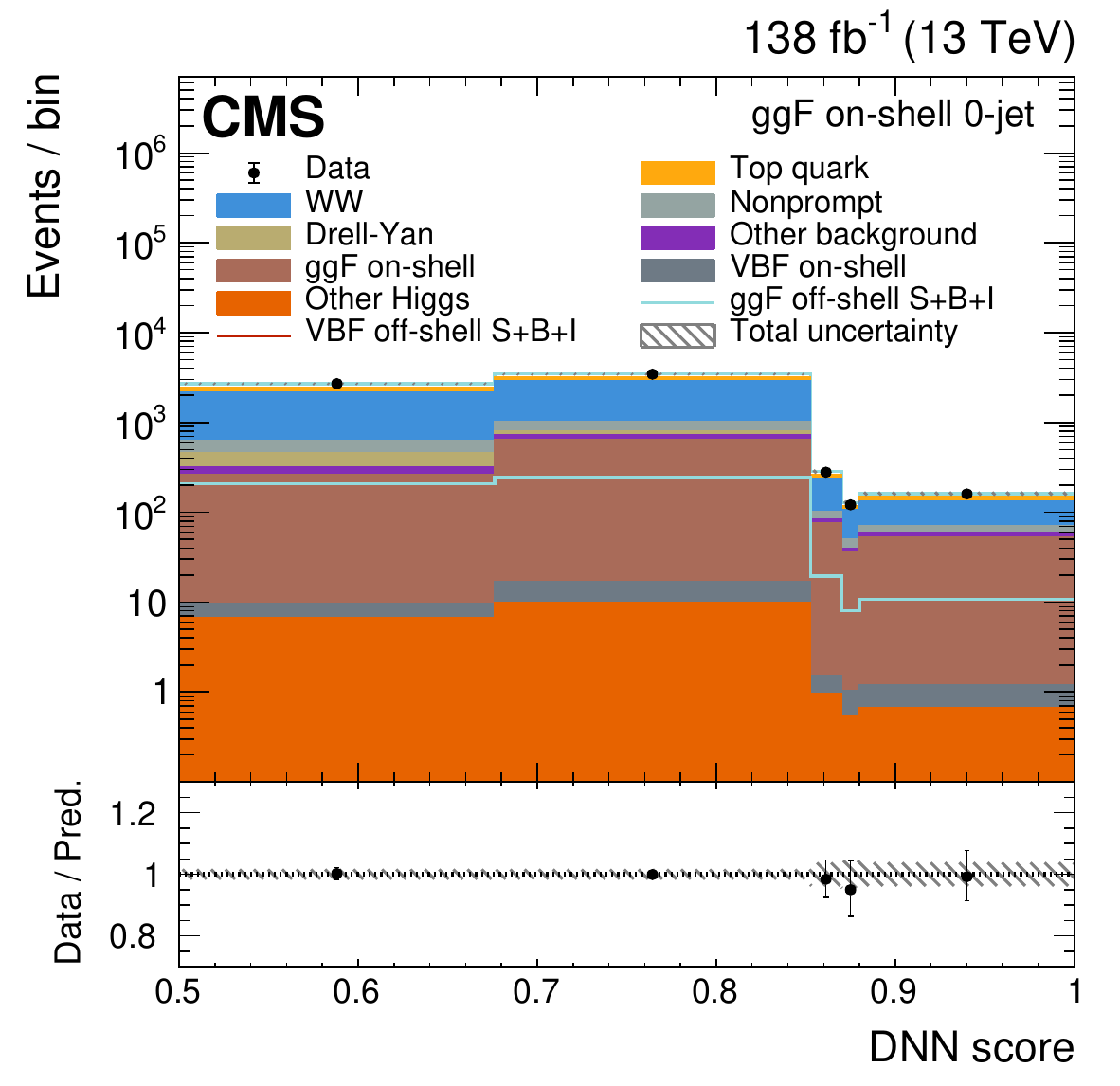}
  \includegraphics[width=0.49\textwidth]{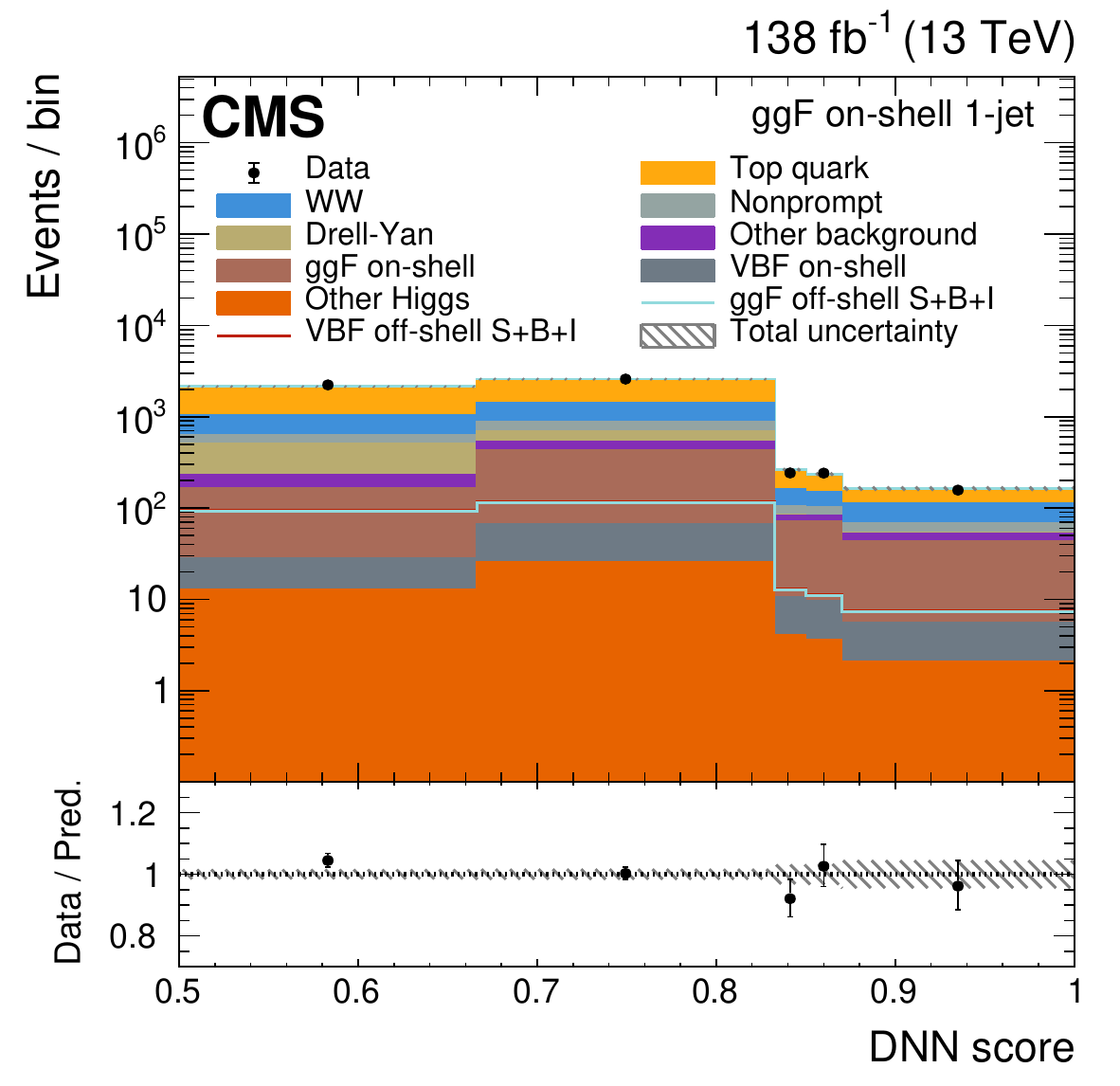} \\
  \includegraphics[width=0.49\textwidth]{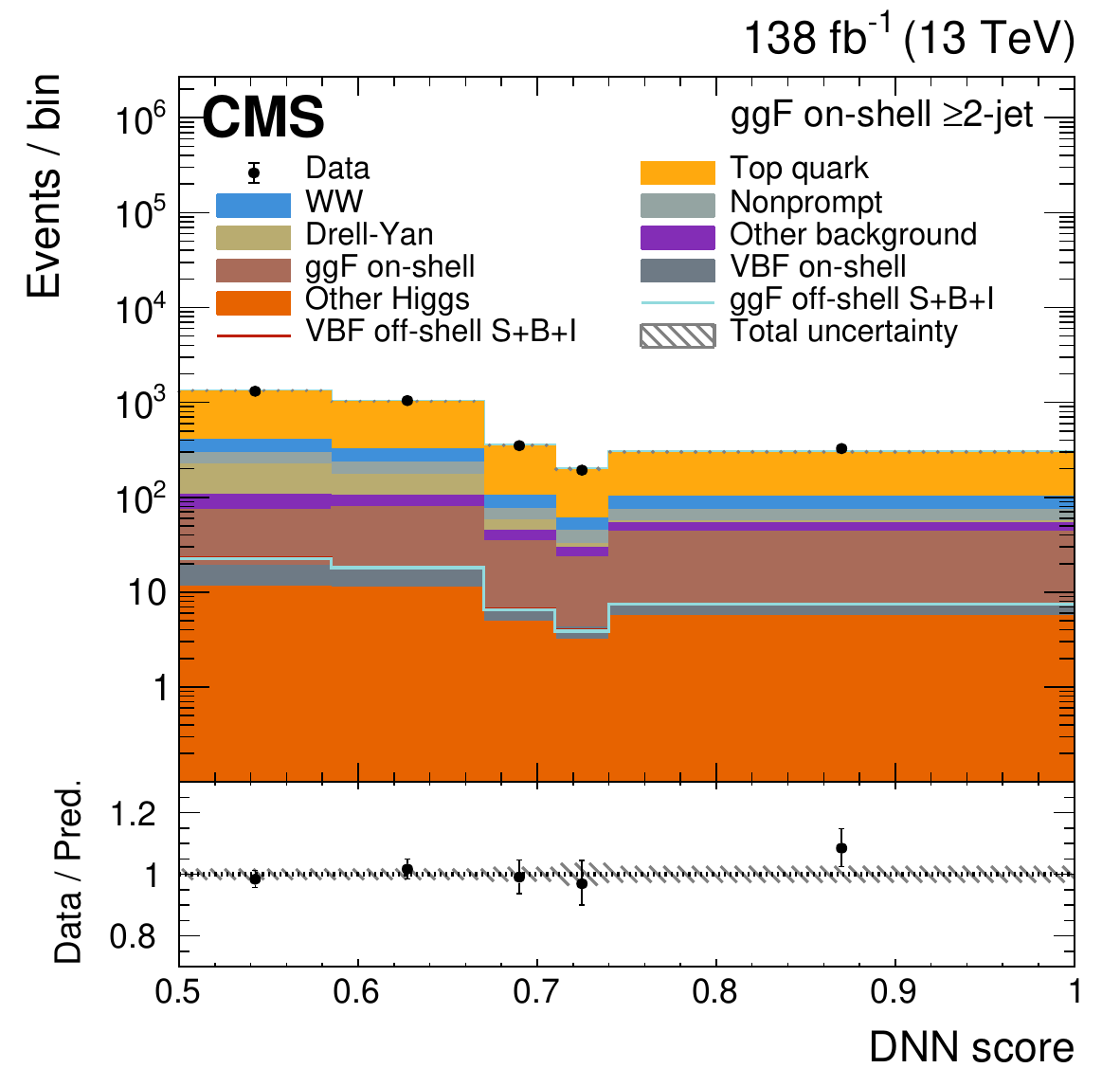}
  \includegraphics[width=0.49\textwidth]{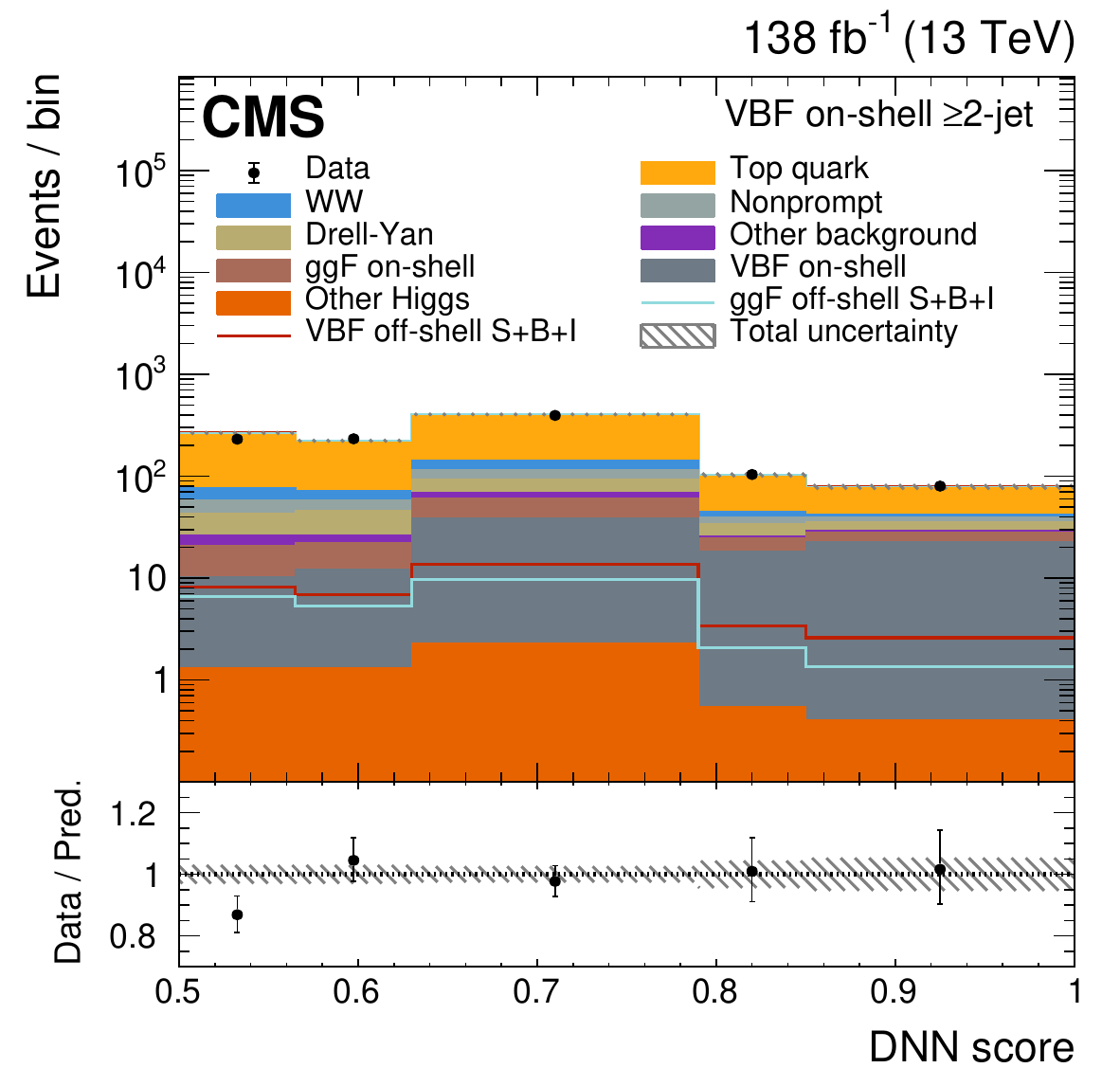}
  \caption{The on-shell CR post-fit DNN output distributions for the 0-jet (upper left), 1-jet (upper right), and $\geq$2-jet (lower plots) event categories. The data are shown as the black dots, with the vertical bars representing the statistical uncertainties. The predicted signal and background distributions are shown by the various colored histograms. The predicted signal ggF and VBF off-shell $S+B+I$ distributions are displayed both as part of the total stacked histograms, along with all the predicted background distributions, and on top of the stacked background distributions for visibility. The lower portion of the plots gives the ratio of the data to the total predicted yields, including the post-fit signal. The hatched area represents the total uncertainty in the ratio.}
  \label{fig:CR_dnnScores}
\end{figure*}

The post-fit distributions for the DNN output in the off-shell SR and on-shell CR categories are shown in Figs.~\ref{fig:post-fit_dnnScores} and~\ref{fig:CR_dnnScores}, respectively. The binning of the distributions has been optimized to maximize the sensitivity in the last three bins, while avoiding large statistical fluctuations. This is accomplished through fine binning, merging bins from right to left until the relative statistical uncertainty in each of the three highest bins is below 0.1. The chi-square goodness-of-fit test with the saturated model~\cite{Cousins2013GeneralizationOC}, used to evaluate how compatible the observed data are with the physics model, returns a p-value~\cite{Gross:2010qma} of 0.073. This value, while relatively low, indicates that the data are statistically consistent with the model at the 95\% CL, and no significant localized tensions are observed.
 
\section{Results}\label{sec:results}
\begin{figure*}[htbp] 
    \centering
    \includegraphics[width=.49\linewidth]{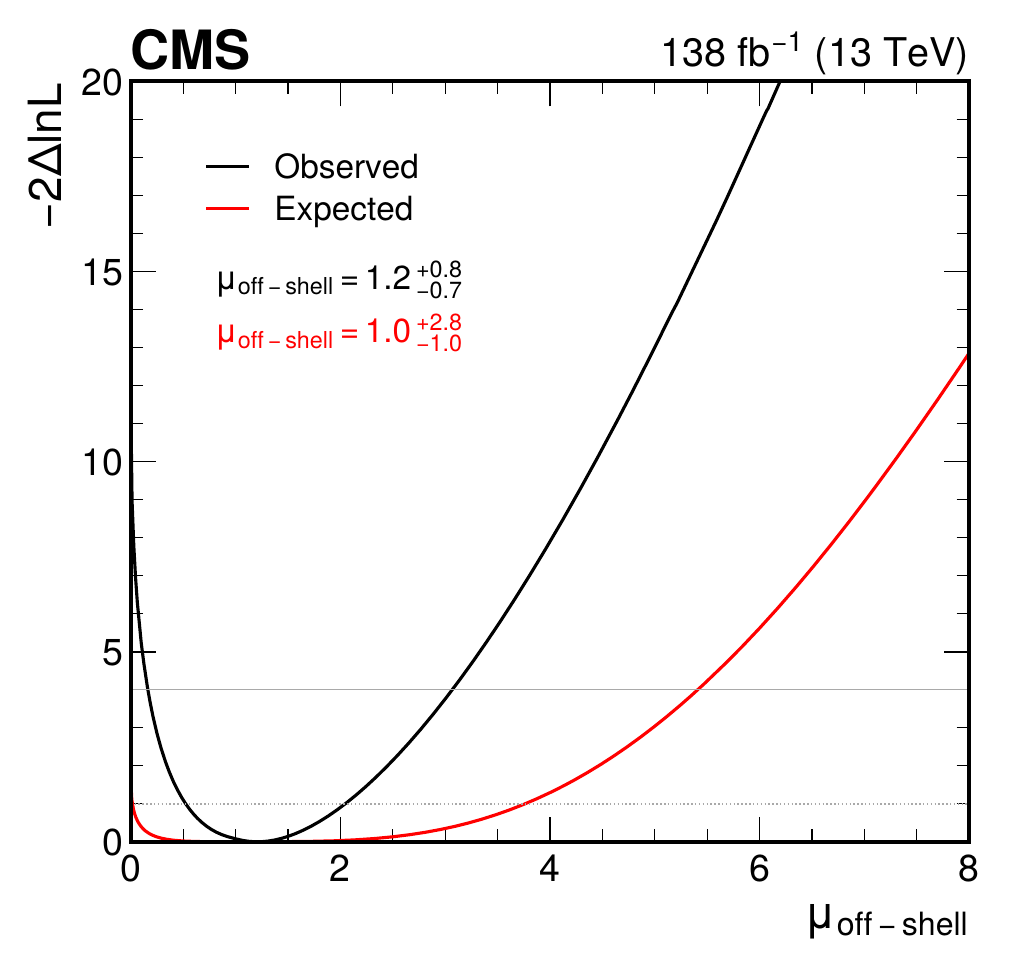}
    \includegraphics[width=.49\linewidth]{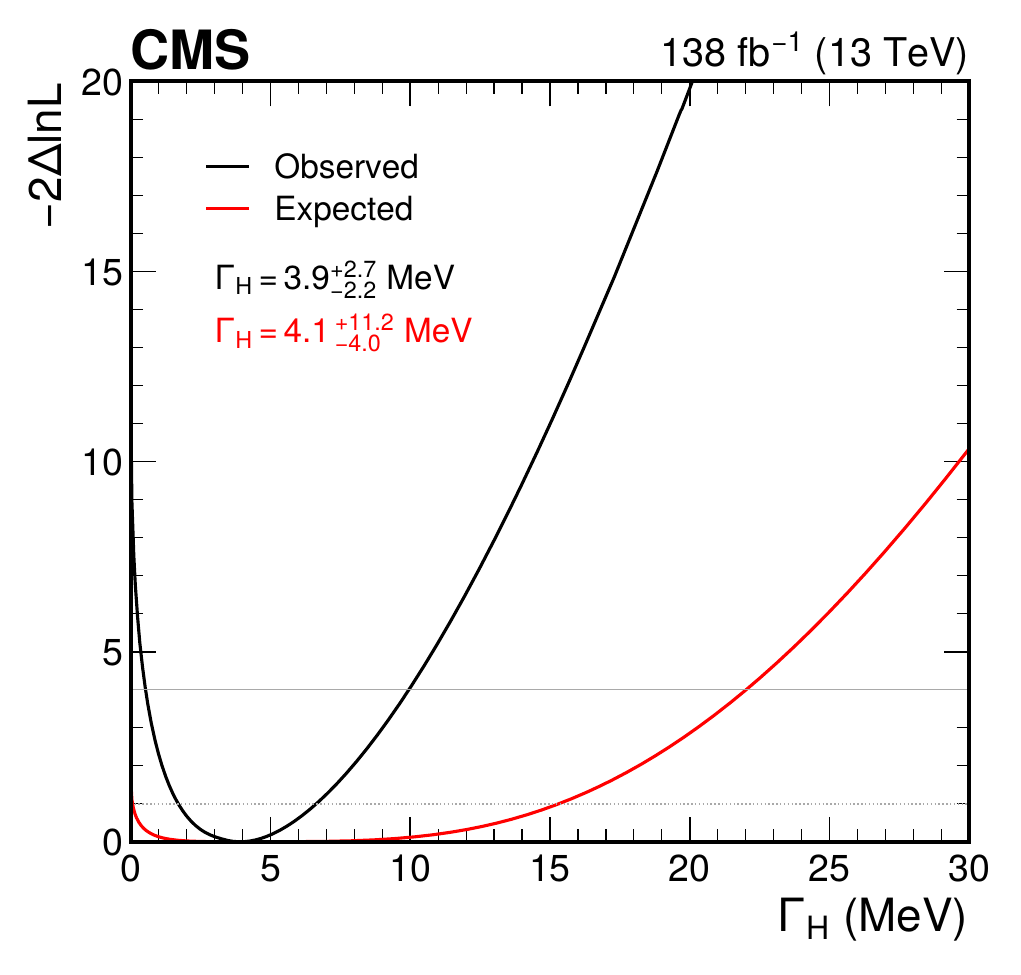}
    \caption{The observed (black curve) and expected (red curve) likelihood function scans for the off-shell signal strength $\mu_{\off}$ (left) and the Higgs boson decay width $\GH$ (right). The resulting best-fit observed and expected values are also given, along with their uncertainties. The 68 and 95\% confidence level intervals are shown by the intersection of the likelihood function curves with the dotted and solid horizontal lines, respectively.
}
    \label{fig:scans_full_run2}  
\end{figure*}

The likelihood scans for the combined off-shell signal strength (left) and the Higgs boson width (right) are shown in Fig.~\ref{fig:scans_full_run2}. 

\begin{table*}[htbp]
    \centering
    \topcaption
    {
      Measurements of the ggF and VBF off-shell signal strengths $\mu^{ggF}_{\off}$ and $\mu^{VBF}_{\off}$, the overall off-shell signal strength $\mu_{\off}$, the on-shell signal strength $\mu_{\on}$, and the Higgs boson decay width $\GH$.
      The observed best-fit values are given (Fit), as well as the observed and expected 68 and 95\% confidence levels (CLs). 
    }
    \renewcommand{\arraystretch}{1.25}
    \begin{scotch}{ccclll} 
    Parameter & Fit & \multicolumn{2}{c}{Observed} & \multicolumn{2}{c}{Expected}  \\
     & & $68\%$~\CL & \multicolumn{1}{c}{$95\%$~\CL} & \multicolumn{1}{c}{$68\%$~\CL} & \multicolumn{1}{c}{$95\%$~\CL} \\
    \hline
    $\mu^{ggF}_{\off}$ & $1.19$ & $0.48, 2.15$ & $\,{<}3.40$ & $\,{<}4.31$ & $\,{<}6.21$ \\
    $\mu^{VBF}_{\off}$ & $1.25$ & $0.16, 3.09$ & $\,{<}5.59$ & $\,{<}6.41$ & $\,{<}11.23$ \\
    $\mu_{\off}$ & $1.21$ & $0.52, 2.05$ & $0.16, 3.08$ & $0.02, 3.75$ & $\,{<}5.41$ \\
    $\mu_{\on}$ & $1.25$ & $1.09, 1.44$ & $0.94, 1.65$ & $0.85, 1.18$ & $0.72, 1.38$ \\
    $\GH$ (\MeVns{}) & $3.91$ & $1.70, 6.61$ & $0.54, 9.94$ & $0.07,  15.25$ & $\,{<}22.00 $ \\
    \end{scotch}
    \label{table:interp}
\end{table*}

Table~\ref{table:interp} gives an overview of the analysis results. The derived value of $\GH$ is $3.9^{+2.7}_{-2.2}\MeV$, where the uncertainties include both the statistical and systematic components, in agreement with the SM prediction of 4.1\MeV. The best fit value for $\mu_{\off}$ is $1.2^{+0.8}_{-0.7}$. The central value of the $\mu_{\off}$ result is larger than 1 even though the highest bins in both plots of Fig.~\ref{fig:post-fit_dnnScores} have a slight deficit below the predicted values. This is because the interference between the off-shell signal and the continuum background is destructive, so a $\mu_{\off}$ value that exceeds unity does not necessarily imply an excess of signal events. This fact contributes to the observed limits on $\mu_{\off}$ being stronger than the expected limits from the background predictions, as shown in Fig.~\ref{fig:scans_full_run2}. The deficits in the signal-enriched DNN output bins favor a scenario with more negative signal--background interference.

The statistical and systematic uncertainties in the ratio of the Higgs boson off-shell to on-shell production discussed in Section~\ref{sec:unc} are summarized in Table~\ref{tab:unc}. It is clear from the table that the total uncertainty in the ratio is dominated by the statistical component. To decompose the total uncertainty in r, nuisance parameters in the model are grouped by categories and likelihood scans are performed by sequentially freezing groups of nuisance parameters and subtracting them in quadrature from the total uncertainty, e.g., the contribution of the statistical uncertainty is obtained by freezing all other nuisance parameters.

\begin{table*}[!htp]
    \centering
    \topcaption{The sources of systematic uncertainty and their contributions to the total uncertainty in the parameter of interest $\text{r} = \mu_{\off}/\mu_{\on}$. The statistical uncertainty is also shown. The values given are the ratios in percent of each uncertainty, divided by the total uncertainty. The last column shows the similar ratios for the uncertainty in $\mu_{\off}$.}
    \begin{scotch}{lcc}
        Uncertainty                         & Contribution to r (\%)  & Contribution to $\mu_{\off}$ (\%)               \\
        \hline
        Integrated luminosity & 2 & 3 \\
        Theoretical & 27 & 29 \\
        Experimental & 8 & 6 \\
        Other & 4 & 3 \\
        Statistical & 56 & 56 \\
    \end{scotch}
    \label{tab:unc}
\end{table*}

The scenario of no off-shell Higgs boson production is excluded at 3.4 standard deviations (1.3 standard deviations expected). This result significantly improves on the previous CMS measurement in the $\PH \to \PW\PW \to 2\Pell2\PGn$ channel at $\sqrt{s} = 7$ and $8\TeV$ of $\GH < 26\MeV$~\cite{CMS_HWWwidth_Run1}, and is comparable with the CMS $\PH \to \PZ\PZ$ $\sqrt{s} = 13\TeV$ measurement of $3.0^{+2.0}_{-1.5}\MeV$~\cite{Hayrapetyan_2025}.

\section{Summary}\label{sec:conclusion}
Measurements of the off- and on-shell signal strengths for the Higgs boson have been performed with the gluon-gluon fusion (ggF) and vector boson fusion (VBF) production modes in the $\PH \to \PW\PW \to \Pe\PGn\PGm\PGn$ decay channel, using proton-proton collision data at $\sqrt{s}=13\TeV$ from the CMS experiment in 2016--2018, corresponding to an integrated luminosity of 138\fbinv. Specific event selection and background discrimination techniques were applied in order to extract the results from maximum likelihood fits to data. The combined off-shell Higgs boson signal strength from ggF and VBF processes is found to be $\mu_{\off} = 1.2^{+0.8}_{-0.7}$, which is used to derive the Higgs boson total decay width of $\GH = 3.9^{+2.7}_{-2.2}\MeV$, in agreement with the standard model prediction of 4.1\MeV. This measurement represents the first CMS determination of the Higgs boson decay width using the $\PH \to \PW\PW$ channel at $\sqrt{s}=13\TeV$. This result improves on the analogous CMS analyses at $\sqrt{s} = 7$ and $8\TeV$ limit by a factor of 3, while complementing the CMS Higgs boson decay width measurement using the $\PH \to \PZ\PZ$ channel at $\sqrt{s}=13\TeV$.

\begin{acknowledgments}
We congratulate our colleagues in the CERN accelerator departments for the excellent performance of the LHC and thank the technical and administrative staffs at CERN and at other CMS institutes for their contributions to the success of the CMS effort. In addition, we gratefully acknowledge the computing centers and personnel of the Worldwide LHC Computing Grid and other centers for delivering so effectively the computing infrastructure essential to our analyses. Finally, we acknowledge the enduring support for the construction and operation of the LHC, CMS detector, and supporting computing infrastructure provided by the following funding agencies: SC (Armenia), BMBWF and FWF (Austria); FNRS and FWO (Belgium); CNPq, CAPES, FAPERJ, FAPERGS, and FAPESP (Brazil); MES and BNSF (Bulgaria); CERN; CAS, MoST, and NSFC (China); MINCIENCIAS (Colombia); MSES and CSF (Croatia); RIF (Cyprus); SENESCYT (Ecuador); ERC PRG, TARISTU24-TK10 and MoER TK202 (Estonia); Academy of Finland, MEC, and HIP (Finland); CEA and CNRS/IN2P3 (France); SRNSF (Georgia); BMFTR, DFG, and HGF (Germany); GSRI (Greece); NKFIH (Hungary); DAE and DST (India); IPM (Iran); SFI (Ireland); INFN (Italy); MSIT and NRF (Republic of Korea); MES (Latvia); LMTLT (Lithuania); MOE and UM (Malaysia); BUAP, CINVESTAV, CONACYT, LNS, SEP, and UASLP-FAI (Mexico); MOS (Montenegro); MBIE (New Zealand); PAEC (Pakistan); MES, NSC, and NAWA (Poland); FCT (Portugal); MESTD (Serbia); MICIU/AEI and PCTI (Spain); MOSTR (Sri Lanka); Swiss Funding Agencies (Switzerland); MST (Taipei); MHESI (Thailand); TUBITAK and TENMAK (T\"{u}rkiye); NASU (Ukraine); STFC (United Kingdom); and DOE and NSF (USA).

\hyphenation{Rachada-pisek} Individuals have received support from the Marie-Curie program and the European Research Council and Horizon 2020 Grant, Contract Nos.\ 675440, 724704, 752730, 758316, 765710, 824093, 101115353, 101002207, 101001205, and COST Action CA16108 (European Union); the Leventis Foundation; the Alfred P.\ Sloan Foundation; the Alexander von Humboldt Foundation; the Science Committee, Project no. 22rl-037 (Armenia); the Fonds pour la Formation \`a la Recherche dans l'Industrie et dans l'Agriculture (FRIA) and Fonds voor Wetenschappelijk Onderzoek Contract No. 1228724N (Belgium); the Beijing Municipal Science \& Technology Commission, No. Z191100007219010, the Fundamental Research Funds for the Central Universities, the Ministry of Science and Technology of China under Grant No. 2023YFA1605804, the Natural Science Foundation of China under Grant No. 12535004, and USTC Research Funds of the Double First-Class Initiative No.\ YD2030002017 (China); the Ministry of Education, Youth and Sports (MEYS) of the Czech Republic; the Shota Rustaveli National Science Foundation, Grant FR-22-985 (Georgia); the Deutsche Forschungsgemeinschaft (DFG), among others, under Germany's Excellence Strategy -- EXC 2121 ``Quantum Universe" -- 390833306, and under Project No. 400140256 - GRK2497; the Hellenic Foundation for Research and Innovation (HFRI), Project No. 2288 (Greece); the Hungarian Academy of Sciences, the New National Excellence Program - \'UNKP, NKFIH Research Grants K 131991, K 133046, K 138136, K 143460, K 143477, K 146913, K 146914, K 147048, 2020-2.2.1-ED-2021-00181, TKP2021-NKTA-64, and 2025-1.1.5-NEMZ\_KI-2025-00004 (Hungary); the Council of Science and Industrial Research, India; ICSC -- National Research Center for High Performance Computing, Big Data and Quantum Computing, FAIR -- Future Artificial Intelligence Research, and CUP I53D23001070006 (Mission 4 Component 1), funded by the NextGenerationEU program (Italy); the Latvian Council of Science; the Ministry of Education and Science, Project No. 2022/WK/14, and the National Science Center, Contracts Opus 2021/41/B/ST2/01369, and 2021/43/B/ST2/01552, 2023/49/B/ST2/03273, and the NAWA contract BPN/PPO/2021/1/00011 (Poland); the Funda\c{c}\~ao para a Ci\^encia e a Tecnologia, Grant CEECIND/01334/2018 (Portugal); the National Priorities Research Program by Qatar National Research Fund; MICIU/AEI/10.13039/501100011033, ERDF/EU, ``European Union NextGenerationEU/PRTR", and Programa Severo Ochoa del Principado de Asturias (Spain); the Chulalongkorn Academic into Its 2nd Century Project Advancement Project, the National Science, Research and Innovation Fund program IND\_FF\_68\_369\_2300\_097, and the Program Management Unit for Human Resources \& Institutional Development, Research and Innovation, Grant B39G680009 (Thailand); the Kavli Foundation; the Nvidia Corporation; the SuperMicro Corporation; the Welch Foundation, Contract C-1845; and the Weston Havens Foundation (USA).
\end{acknowledgments}

\bibliography{auto_generated}
\cleardoublepage \appendix\section{The CMS Collaboration \label{app:collab}}\begin{sloppypar}\hyphenpenalty=5000\widowpenalty=500\clubpenalty=5000
\cmsinstitute{Yerevan Physics Institute, Yerevan, Armenia}
{\tolerance=6000
A.~Hayrapetyan, V.~Makarenko\cmsorcid{0000-0002-8406-8605}, A.~Tumasyan\cmsAuthorMark{1}\cmsorcid{0009-0000-0684-6742}
\par}
\cmsinstitute{Institut f\"{u}r Hochenergiephysik, Vienna, Austria}
{\tolerance=6000
W.~Adam\cmsorcid{0000-0001-9099-4341}, L.~Benato\cmsorcid{0000-0001-5135-7489}, T.~Bergauer\cmsorcid{0000-0002-5786-0293}, M.~Dragicevic\cmsorcid{0000-0003-1967-6783}, P.S.~Hussain\cmsorcid{0000-0002-4825-5278}, M.~Jeitler\cmsAuthorMark{2}\cmsorcid{0000-0002-5141-9560}, N.~Krammer\cmsorcid{0000-0002-0548-0985}, A.~Li\cmsorcid{0000-0002-4547-116X}, D.~Liko\cmsorcid{0000-0002-3380-473X}, M.~Matthewman, J.~Schieck\cmsAuthorMark{2}\cmsorcid{0000-0002-1058-8093}, R.~Sch\"{o}fbeck\cmsAuthorMark{2}\cmsorcid{0000-0002-2332-8784}, M.~Shooshtari\cmsorcid{0009-0004-8882-4887}, M.~Sonawane\cmsorcid{0000-0003-0510-7010}, W.~Waltenberger\cmsorcid{0000-0002-6215-7228}, C.-E.~Wulz\cmsAuthorMark{2}\cmsorcid{0000-0001-9226-5812}
% ...
\par}
\vskip\cmsinstskip
\dag:~Deceased\\
$^{1}$Also at Yerevan State University, Yerevan, Armenia\\
$^{2}$Also at TU Wien, Vienna, Austria\\
% ...
\end{sloppypar}
\end{document}